\documentclass[%
reprint,
 amsmath,amssymb,
 aps,
 prb, 
]{revtex4-1}

\usepackage{graphicx}% Include figure files
\usepackage{dcolumn}% Align table columns on decimal point
\usepackage{bm}% bold math
\usepackage{color}

\begin{document}

\preprint{APS/123-QED}

\title{Quantum confinement effects in Pb Nanocrystals grown on InAs}
\author{Tianzhen Zhang}
\affiliation{LPEM, ESPCI Paris, PSL Research University, CNRS, Sorbonne Universit\'e, 75005 Paris, France}

\author{Sergio Vlaic}
\affiliation{LPEM, ESPCI Paris, PSL Research University, CNRS, Sorbonne Universit\'e, 75005 Paris, France}

\author{St\'ephane Pons}
\affiliation{LPEM, ESPCI Paris, PSL Research University, CNRS, Sorbonne Universit\'e, 75005 Paris, France}

\author{Guy Allan}
\affiliation{Univ. Lille, CNRS, Centrale Lille, ISEN, Univ. Valenciennes, UMR 8520 -
IEMN, F-59000 Lille, France}

\author{Christophe Delerue}
\affiliation{Univ. Lille, CNRS, Centrale Lille, ISEN, Univ. Valenciennes, UMR 8520 -
IEMN, F-59000 Lille, France}

\author{Alexandre Assouline}
\affiliation{LPEM, ESPCI Paris, PSL Research University, CNRS, Sorbonne Universit\'e, 75005 Paris, France}

\author{Alexandre Zimmers}
\affiliation{LPEM, ESPCI Paris, PSL Research University, CNRS, Sorbonne Universit\'e, 75005 Paris, France}

\author{Christophe David}
\affiliation{Centre de Nanosciences et de Nanotechnologies, CNRS, Univ. Paris-Sud, Universit\'es Paris-Saclay, C2N – Marcoussis, 91460 Marcoussis, France}

\author{Guillemin Rodary}
\affiliation{Centre de Nanosciences et de Nanotechnologies, CNRS, Univ. Paris-Sud, Universit\'es Paris-Saclay, C2N – Marcoussis, 91460 Marcoussis, France}

\author{Jean-Christophe Girard}
\affiliation{Centre de Nanosciences et de Nanotechnologies, CNRS, Univ. Paris-Sud, Universit\'es Paris-Saclay, C2N – Marcoussis, 91460 Marcoussis, France}

\author{Dimitri Roditchev}
\affiliation{LPEM, ESPCI Paris, PSL Research University, CNRS, Sorbonne Universit\'e, 75005 Paris, France}

\author{Herv\'e Aubin}
\email{Herve.Aubin@espci.fr} 
\affiliation{LPEM, ESPCI Paris, PSL Research University, CNRS, Sorbonne Universit\'e, 75005 Paris, France}

\date{\today}% It is always \today, today,

\begin{abstract}
In the recent work of Ref.\cite{Vlaic2017-bs}, it has been shown that Pb nanocrystals grown on the electron accumulation layer at the (110) surface of InAs are in the regime of Coulomb blockade. This enabled the first scanning tunneling spectroscopy study of the superconducting parity effect across the Anderson limit. The nature of the tunnel barrier between the nanocrystals and the substrate has been attributed to a quantum constriction of the electronic wave-function at the interface due to the large Fermi wavelength of the electron accumulation layer in InAs. In this manuscript, we detail and review the arguments leading to this conclusion. Furthermore, we show that, thanks to this highly clean tunnel barrier, this system is remarkably suited for the study of discrete electronic levels induced by quantum confinement effects in the Pb nanocrystals. We identified three distinct regimes of quantum confinement. For the largest nanocrystals, quantum confinement effects appear through the formation of quantum well states regularly organized in energy and in space. For the smallest nanocrystals, only atomic-like electronic levels separated by a large energy scale are observed. Finally, in the intermediate size regime, discrete electronic levels associated to electronic wave-functions with a random spatial structure are observed, as expected from Random Matrix Theory.
\end{abstract}

\pacs{Valid PACS appear here}% PACS, the Physics and Astronomy
\maketitle

\section{Introduction}
Atomic clusters and NanoCrystals (NCs) offer promising perspectives for the study of electronic orders and correlations at the scale of single electronic states in the quantum confined regime, through statistical analysis of the energy levels distribution or the study of the spatial structures of the corresponding wave-functions.

The quantum confined regime is reached when the electronic level spacing $\delta$ in a NC is larger than the thermal broadening $k_{\rm B}T$ and larger than the coupling $\Gamma$ of these levels to the continuum of states within the metallic electrodes employed for measurements. In this regime, the energy of the discrete levels and the spatial distribution of the associated wave-functions depend on the boundary conditions imposed by the surface of the NC.

The quantum confined regime has been intensively studied in semiconducting Quantum Dots (QDots), either in colloidal QDots\cite{Klein1997-gb,Banin1999-oj, Millo2000-jr, Millo2001-ls,Liljeroth2005-mw, Liljeroth2006-ge, Liljeroth2006-iy, Jdira2006-er, Sun2009-ph,Wang2015-kh,Wang2017-su} or in micro-fabricated QDots\cite{Chang1996-mw,Folk1996-xi,Cronenwett1997-ax,Patel1998-qg,Huibers1998-mh,Patel1998-oo}, where the mean level spacing: 
\begin{equation}
\label{levelspacing}
\langle\delta\rangle=\frac{2(\pi\hbar)^2}{m^* k_{\rm F} \textup{Volume}}
\end{equation} is large because of the large Fermi wavelength $\lambda_{\rm F}=2\pi/k_{\rm F}$, which makes the experimental identification of the discrete levels in the spectrum easier than in metallic QDots where the Fermi wavelength is short.

The energy distribution of the electronic levels and the structure of the associated wave-functions is expected to depend on the location of the energy levels with respect to the Thouless energy $E_{\rm T}=h/\tau$ where $\tau$ is the propagation time of the electrons across the QDot. With this definition of the Thouless energy, one finds that the ratio $E_{\rm T}/\langle\delta\rangle=(4\pi/3)(r/\lambda_{\rm F})^2\simeq 0.5(r/\lambda_{\rm F})^2$ depends only on the ratio of the QDot radius r with the Fermi wavelength.
Two regimes of quantum confinement can be distinguished with $E_{\rm T}/\langle\delta\rangle >1$ for the chaotic regime and $E_{\rm T}/\langle\delta\rangle < 1$ for the regular regime.

The first regime, where the level spacing is smaller than the Thouless energy, is reached in QDots of radius much larger than the Fermi wavelength. In this case, for electronic states whose energy measured with respect to the Fermi energy is smaller than the Thouless energy, the electronic wave-functions are delocalized on the whole QDot and consequently, the electronic states are correlated through their Fermi statistics. In presence of electron-electron interactions and disorder, this leads to the formation of a complex quantum system with chaotic dynamic\cite{Bohigas1984-yw,Haake2001-fu}. The energy levels are expected to follow a distribution $P(\varepsilon)$ given by Random Matrix Theory (RMT)\cite{Alhassid2000-cd,Mirlin2000-xa} characterized by levels repulsion $P(\varepsilon\rightarrow 0)\simeq 0$ at zero energy as a consequence of Pauli exclusion. Furthermore, the wave-functions are expected to display large amplitude fluctuations with random or possibly fractal structure\cite{Jalabert1992-dx}. This chaotic regime has been studied in micrometer sized QDots micro-fabricated from III-V heterostructures and measured at milli-kelvin temperature in dilution fridges\cite{Chang1996-mw,Folk1996-xi,Cronenwett1997-ax,Patel1998-qg,Huibers1998-mh,Patel1998-oo}. Fluctuations of the amplitude of the wave-function  have been observed to govern the statistics  of  the  tunnel  conductance  in  the  Coulomb blockade regime\cite{Jalabert1992-dx,Chang1996-mw,Huibers1998-mh,Assouline2017-nx}. 

%Furthermore, electronic states of energy higher than the Thouless energy are localized and the energy levels follow a Poisson distribution.

%RMT was first developed by Wigner\cite{Wigner1951-zy}, then by Dyson\cite{Dyson1962-fr} and Mehta\cite{Mehta2004-lq} in the early 1960s. It is now believed that the RMT describes adequately statistical properties of spectra of quantum systems whose classical analogs are chaotic\cite{Bohigas1984-yw}. In particular, employing a supersymmetric method, it was demonstrated by Efetov that RMT describes properly the level statistics of disordered systems, which is also relevant for to the theoretical description of Anderson metal-insulator transitions\cite{Evers2008-jz}.

The second regime, where the level spacing is larger than the Thouless energy,  is reached in QDots of radius smaller than the Fermi wavelength or, equivalently, the Bohr radius. In that case, no level repulsion and no chaotic regime is expected. The electronic states follow a regular pattern described by atomic-like quantum numbers 1S, 1P and so on. This regime has been mostly studied through tunneling spectroscopy on colloidal QDots by Scanning Tunnel Microscopy (STM) \cite{Banin1999-oj, Millo2000-jr, Millo2001-ls,Liljeroth2005-mw, Liljeroth2006-ge, Liljeroth2006-iy, Jdira2006-er, Sun2009-ph} or on-chip tunnel spectroscopy\cite{Klein1997-gb,Wang2015-kh,Wang2017-su}. STM mapping of the electronic wave-functions has been attempted on colloidal QDots of InAs\cite{Millo2001-ls}. As expected in those nanometer-sized QDots of radius smaller than the Bohr radius, the wave-functions had simple spherical and toroidal structures expected for the S and P symmetry respectively.

Mapping of the random or fractal structure of wave-functions of QDots in the chaotic regime has not been achieved so far. As described above, this chaotic regime can be reached in large micrometer-sized QDots; however, STM studies of these micro-fabricated QDots is not possible. The chaotic regime can also be reached in nanometer-sized NCs of high carrier density such as metallic NCs where the Fermi wavelength is very small. In addition to level repulsion, more spectacular electronic correlations effects and electronic orders, such as superconductivity, can also be expected in NCs of high carrier density.

The study of quantum confinement effects in high carrier density materials such as metallic and superconducting NCs is, however, challenging because of the short Fermi wavelength. For Pb, which has a Fermi energy $E_{\rm F}=9.4$~eV, two Fermi surfaces FS$_1$ and FS$_2$, of characteristics wave-vectors $k_{\rm F1}$=7.01 nm$^{-1}$, $k_{\rm F2}$=11.21 nm$^{-1}$ and effective mass m$^*$=1.2 m$_e$; one finds that at the temperature $T=1.3~$K, i.e. thermal broadening $k_{\rm B}T=112~\mu$eV, a discrete electronic spectrum is expected for NCs volume smaller than 650~nm$^3$, i.e. a sphere or radius 5.5~nm. 

In this nanometer size range, discrete electronic levels in metallic grains have been observed through tunneling spectroscopy of nano-fabricated drain-source structures with evaporated aluminum grains\cite{Ralph1999-uh,Ralph1997-ab,Black1996-cd} or chemically synthesized gold nanoparticles\cite{Kuemmeth2008-vg}.

STM is particularly well adapted to the study of quantum confinement effects in NCs. The instrument can provide both a topographic image of the NCs and spectroscopic data with atomic resolution. This should allow not only the observation of discrete electronic levels but also a mapping of the corresponding wave-functions.

While numerous works exist on the STM study of Ultra-High Vacuum (UHV) grown metallic clusters strongly coupled to the substrate\cite{Brune1998-eq}, the study of quantum confinement effects in isolated UHV grown NCs has been hampered by two contradicting requirements: on the one hand, the substrate should be conducting enough to enable a current path to the ground; on the other hand, the NCs should also be separated from this substrate by a second tunneling barrier to preserve quantum confinement, that is the coupling $\Gamma$ between the substrate and the NC should be smaller than the level spacing, i.e. $\Gamma < \delta$.

To enable the growth of high quality crystalline NCs, this second tunnel barrier should present a well-ordered surface. While tunnel barriers on a metallic substrate can be formed by oxidation or the deposition of a dielectric\cite{Bose2010-ob,Hong2013-hm}, these \emph{extrinsic} tunnel barriers usually present poor atomic orders not suitable for the growth of highly crystalline NCs. To avoid the use of these extrinsic tunnel barriers, it should be possible to grow metallic NCs on semiconductor substrates to make use of the Schottky barrier as a tunnel barrier. Unfortunately, so far, most published works\cite{Feenstra1989-rq,McLean1989-dw,First1989-xd} have shown that the Schottky barriers are usually too opaque to enable electron tunneling at low energy, $<$1~eV.

As first described in Ref.~\cite{Vlaic2017-bs}, we discovered another type of \emph{intrinsic} tunnel barrier on the (110) surface of InAs, suitable for the study of quantum confinement in nano-sized metallic NCs and enabling tunneling spectroscopy even at very low energy. The (110) surface of InAs presents an electron accumulation layer whose Fermi wavelength is about 20~nm. When metallic NCs are grown on this surface, their electronic coupling with the electron accumulation layer is strongly reduced when their lateral size is smaller than this Fermi wavelength. This phenomenon has been extensively employed in quantum point contacts where tunnel barriers are obtained by squeezing the electron gas with top split-gates to a distance smaller than the Fermi wavelength\cite{Van_Wees1988-aj,Pasquier1993-pi}.

In the first section of the manuscript, we review the experimental data and the arguments leading to the identification of this new type of intrinsic tunnel barrier. In the second section, we describe our studies of quantum confinement in Pb NCs. We identified three distinct regimes of quantum confinement. In the largest NCs of volume about 100-200~nm$^3$, one observes a Fabry-Perot regime characterized by discrete quantum levels with regular spatial structure and energy distribution. In the smallest NCs of volume about 1~nm$^3$, i.e. atomic clusters, one observes an atomic-like regime characterized by well defined and well separated atomic-like levels. Finally, in NCs of intermediate volume about 10~nm$^3$, one finds discrete electronic levels where the corresponding wave-functions have a random structure. This last regime corresponds to the chaotic regime as described by RMT. These observations confirm the system Pb NCs on InAs as the most suitable for the study of quantum confinement effects in metallic/superconducting NCs.

%As the volume of the unit cell of Pb is V_{\rm cell}=0.12~nm$^3$, and each cell provides eight levels to the electron band, a discrete electronic spectrum for NC volume smaller than 83000~V$_{\rm cell}$/8$\simeq$1200~nm$^3$, a sphere or diameter 28 nm. 

\section{A new type of tunnel barrier}

\subsection{Nanocrystal growth}
The NCs are grown on the (110) surface of InAs, for which an STM image with atomic resolution is shown Fig.~1a. This surface is obtained by cleaving an $\langle 001 \rangle$ oriented substrate, which is n-doped with sulfur to a carrier concentration of N$_D \simeq 6\times 10^{16}$~cm$^{-3}$. The Pb NCs are obtained by thermal evaporation of a nominal 0.3 monolayer of Pb on the substrate heated at T=150$~^\circ$C.

\begin{figure}[h!]
	\begin{center}
		\includegraphics[width=8cm]{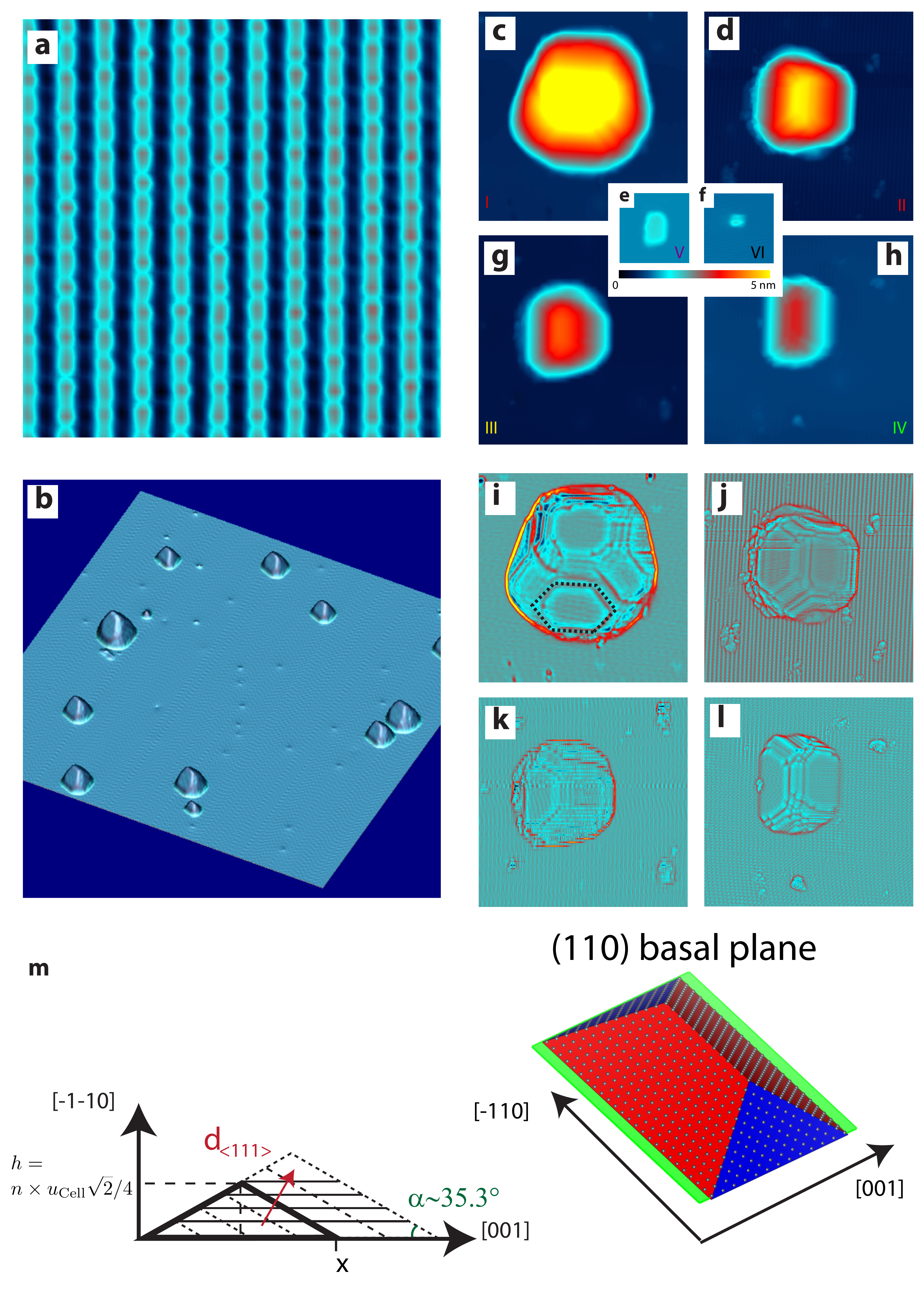}
		\caption{\label{Fig1} a) 6.5~nm$\times$6.5~nm atomic resolution image of InAs (110)(1 V, 30 pA). b) 3D 150~nm $\times$ 150~nm topographic STM image (1 V, 30 pA) of Pb NCs grown on the (110) InAs surface. c) to h) 6 topographic images of NCs of different sizes shown with the same x,y and z scales. c), d) g) and h) are 30~nm$\times$30~nm, while e) and f) are 10~nm$\times$10~nm. i) to l) are Laplacian  $\Delta_{xy}z(x,y)$ images of NCs, corresponding to the topographic images c), d), g) and h). The hexagonal dash line on panel i highlights the hexagonal shape of the facet of the NC. m) Sketch of the pyramidal NC indicating the main crystallographic directions.}
	\end{center}
\end{figure}

As shown on the topographic images, Fig.~1, Pb grows in the Volmer-Weber, i.e. island mode\cite{Brune1998-eq}. Island growth of metals evaporated on III-V semiconductor has been observed for various elements on GaAs such as Ag\cite{Bolmont1982-aa,Ludeke1983-iz,Trafas1991-bi,Neuhold1997-jv}, Au\cite{Feenstra1989-rq}, Fe\cite{First1989-xd} as well as metals on InAs such as Co\cite{Wiebe2003-ry,Muzychenko2009-rc}. In this manuscript, we will focus on the six NCs shown Fig.~1c-h. The corresponding differential images $\nabla_{xy}z$, Fig.~1i-l, for the 4 largest NCs show that the islands are well crystallized and expose mostly the (111) planes of the cubic face-centered Pb structure, as indicated by the observation of the characteristic hexagonal shape of the (111) facets. Surrounding these NCs, the surface remains free from any adsorbate and atomic resolution on the (110) InAs surface is possible.

The shape of the NCs is mostly pyramidal as sketched Fig.~1m, where the [110] direction of Pb is oriented perpendicular to the substrate. For this geometry, the height of the nanocrystal is given by $h=n\times u_{\rm cell}\times\sqrt{2}/4$ where $u_{\rm cell}$=0.495 nm is the length of the unit cell of Pb; $u_{\rm cell}\sqrt{2}/4$ is the distance between atomic rows along the [110] direction of Pb and n is the number of atomic rows. See Supp. Info.\cite{SuppInfo} for more details on the determination of crystallographic orientations.

%The interface between metals and GaAs usually present a large Schottky barrier that prevents tunneling at low energies ($< 1$~eV) \cite{Feenstra1989-rq,McLean1989-dw,First1989-xd,Suzuki1991-ul,Reusch2004-pu,Ifflander2015-tl}.

\subsection{Metal-Semiconductor Interfaces}

To clearly understand the nature of the Pb/InAs interface and the peculiar nature of the tunnel barrier requires, first, to review the current understanding of metal/semiconductor interfaces; see Ref.\cite{Monch2003-fu} for a detailed review.

\begin{figure*}[ht!]
	\begin{center}
		\includegraphics[width=11cm]{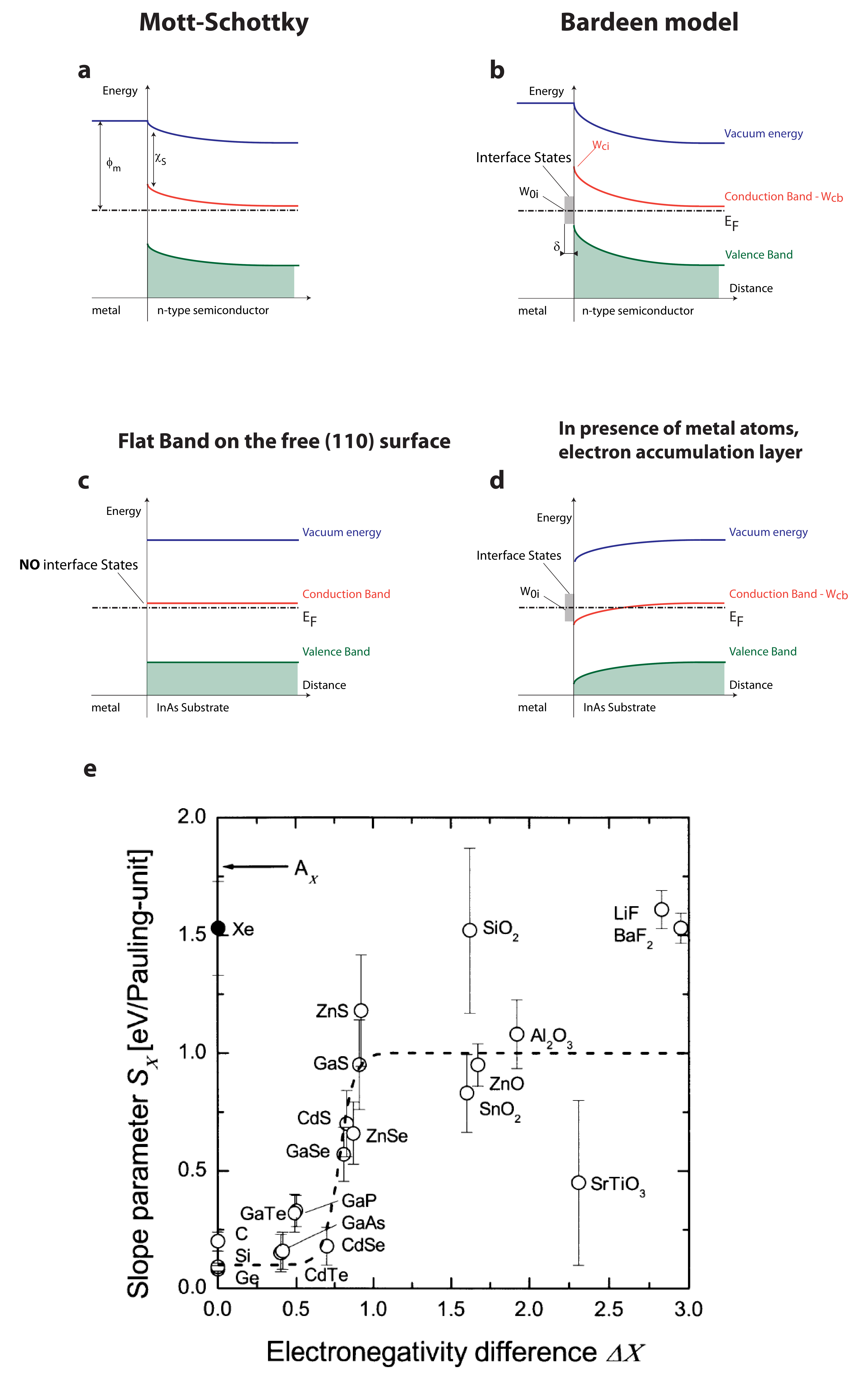}
		\caption{\label{Fig2} a) Mott-Schottky model of band-bending, which applies in the \emph{absence} of interface states. b) Bardeen model of band-bending, which applies in the \emph{presence} of interface states. c) The (110) surface of InAs cleaved in UHV is free of interface states, implying that the Fermi level is not pinned and the bands remain flat up to the surface. d) Because the charge neutrality level $W_{\rm 0i}$ of InAs is above the conduction band, the deposition of a metal layer on the top of InAs leads to interface states and an accumulation layer of electron below the metallic layer. e) The slope parameter $S_{\phi}$ as a function of electronegativity difference for a series of binary semiconductors extracted from Ref.~\cite{Monch2003-fu}. Ionic semiconductors, i.e. high electronegativity difference, tend to have a large slope parameter, which implies that the Mott-Schottky model applies. In contrast, covalent semiconductors such as GaAs and InAs (not shown on the graph) tend to have a small slope parameter implying the presence of a large density of interface states that pin the Fermi level at the charge neutrality level, i.e. Bardeen model.
		}
	\end{center}
\end{figure*}

On a semiconductor surface, the adsorption of a metal overlayer leads to band-bending as sketched Fig.~2ab. The barrier height $\phi_{\rm Bn}$ for the electrons in the conduction band is given by the relation :
\begin{equation}\label{Eq:barrier}
\phi_{\rm Bn}=S_{\phi}(\phi_{\rm m} -\chi_{\rm s})+(1-S_{\phi})(W_{\rm ci}-W_{\rm 0i})
\end{equation}

where $\phi_{\rm m}$ is the metal work function, $\chi_{\rm s}$ is the semiconductor electron affinity, $W_{\rm ci}$ is the energy of the bottom of the conduction band, $W_{\rm 0i}$ is the charge-neutrality level and $S_{\phi}=\frac{\partial \phi_{\rm Bn}}{ \partial \phi_{\rm m}}$ is the slope parameter. This last parameter describes the dependence of the barrier height on the metal work-function. When no interface states are present, the barrier height is given by the well-known Schottky-Mott formula, $\phi_{\rm Bn}=\phi_{\rm m} -\chi_{\rm s}$, i.e. $S_{\phi}=1$ in Eq.~\ref{Eq:barrier}. In this case, the barrier height is proportional to the amplitude of the metal work function. In the other limit, when interface states are present, the barrier height is given by the Bardeen formula, $\phi_{\rm Bn}=W_{\rm ci}-W_{\rm 0i}$, i.e. $S_{\phi}=0$ in Eq.~\ref{Eq:barrier}. In this case, the Fermi level is pinned by interface states at the charge neutrality level. The charge-neutrality level separates the electron-type levels from the hole-type levels. As described in Refs.\cite{Monch2003-fu,Monch2001-ab,Monch1990-nr,Louie1976-lw}, because the interface states derive from the band structure of the semiconductor, the charge neutrality level is an intrinsic property of the semiconductor, which implies that the barrier height does not depend on the metal work function.

%More generally, the slope parameter can be described by the relation :
%\begin{equation}
%S_{\phi}={\left ( {1+{\frac {{e}^{2}} {{{\varepsilon }_{\rm i}}{{\varepsilon }_{0}}}{{D}_{\rm is}{{\delta }_{\rm i}}}}} \right )}^{-1}
%\end{equation}

%where $\varepsilon_{\rm i}$ is the dielectric coefficient at the interface, $D_{\rm is}$ is the density of interface states, $\delta _{\rm i}$ is the thickness of the dipole layer at the interface.

A plot of the slope parameter measured experimentally for many distinct semiconductors, Fig.2e, shows that in III-V semiconductors the slope parameter is close to zero, indicating that the barrier height is set by the Bardeen relation, where a large density of interface states is induced by the metal overlayer and leads to a pinning of the Fermi level.

For most III-V semiconductors, the charge neutrality level lies within the band gap except for small gap materials like InAs and InSb, where the charge neutrality level has been found within the conduction band\cite{Tersoff1984-ow, Tersoff1984-gd,Monch2001-ab,Morgenstern2012-oa}. Consequently, the adsorption of a metal overlayer leads to the formation of an accumulation layer of electrons, as sketched Fig.~2d. In particular, on InAs (110), the Fermi energy is found at about 100-400 meV above the conduction band minimum upon deposition of different adsorbates such as H, O, N, Cl, Ag, Au, Ga, Cu, Cs, Na, Sb, Nb, Fe and Co\cite{Baier1986-ua,Chen1989-jf,Smit1989-uk,Aristov1991-ms,Van_Gemmeren1993-om,Aristov1993-bp,Aristov1993-bp,Aristov1994-io,Aristov1994-hs,Aristov1995-ut,Nowak1995-sr,Morgenstern2000-ig, Getzlaff2001-ii, Morgenstern2002-pu}. For Pb, while no data exists for the (110) surface, it was shown that one mono-layer of Pb on the (100) surface of InAs leads to an accumulation layer of electrons\cite{Layet1998-ci}. 

%Early STM works\cite{First1989-xd} have identified metal-induced gap states near Fe clusters deposited on GaAs.

Pinning of the Fermi level can also occur at bare UHV cleaved surfaces as consequence of native surface defects. On InAs, this leads to the formation of an accumulation layer at (100) and (111) surfaces\cite{Olsson1996-vm,Piper2006-is}, but not on the (110) surface where the band remains flat, Fig.~2c. This results from the fact that the (110) surface contains the same number of cations and anions per unit area. Thus, this surface is intrinsically neutral,  free from defects and no reconstruction is ever observed for this surface\cite{Monch2001-ab}. Consequently, no pinning of the Fermi level is ever observed for the UHV cleaved (110) surface.

One remarkable consequence of the absence of Fermi level pinning is the possibility to shift the conduction band energy with the electric field induced by the STM tip. This leads to the formation of a QDot induced by the band bending generated by the STM tip, as sketched Fig.~3a.

\subsection{Tip-induced QDot Levels}

\begin{figure}[ht!]
	\begin{center}
		\includegraphics[width=8cm]{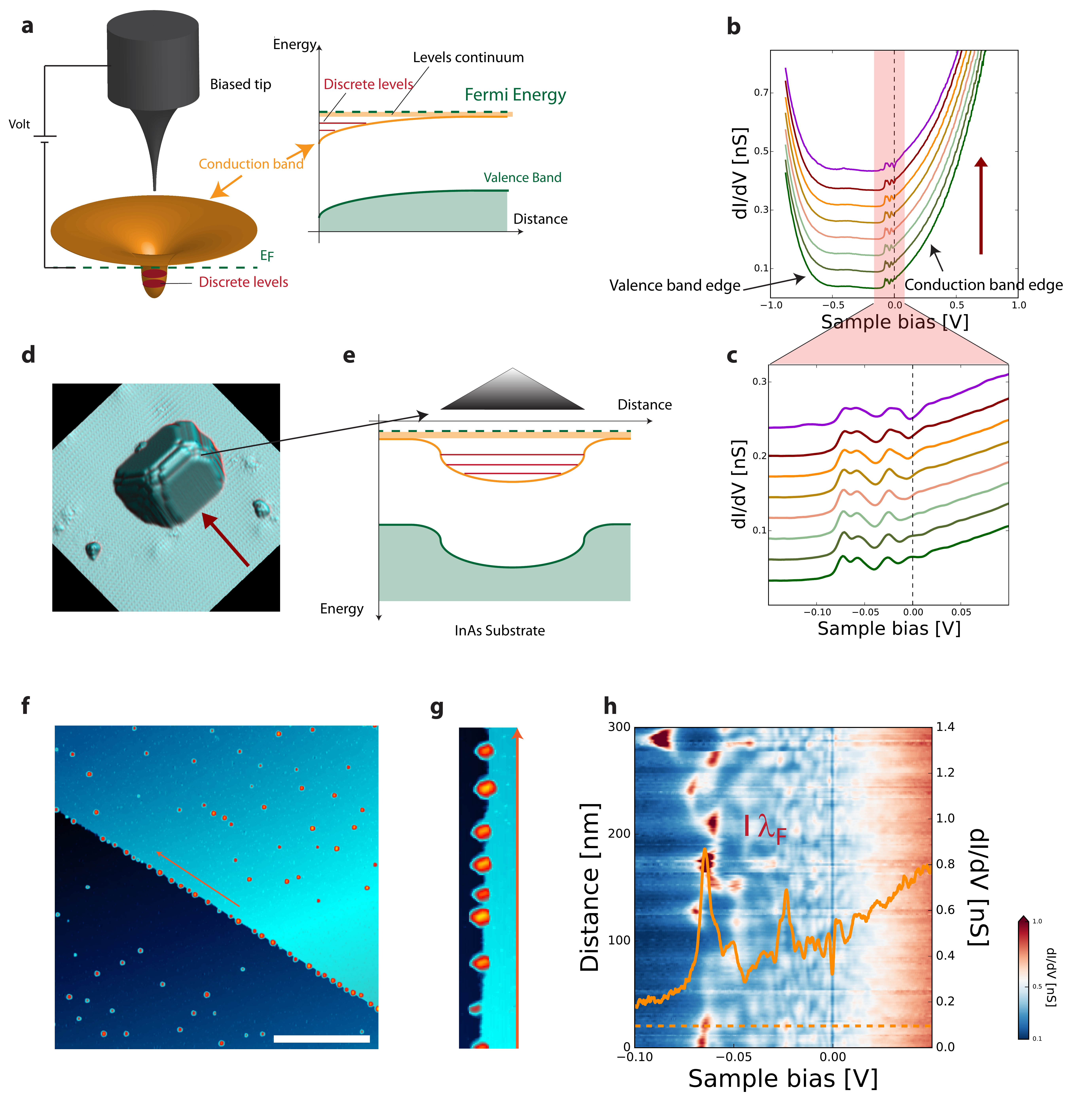}
		\caption{\label{Fig_TQDOT} a) The band bending induced by the tip leads to the formation of a QDot into the accumulation layer at the surface of InAs. This QDot has discrete electronic levels that are seen  as conductance peaks into  the tunneling spectra shown panels b) c) and h). b) Conductance spectrum dI/dV measured along the red arrow shown panel d. The edges of the conduction and valence bands are clearly visible, as well as the conductance peaks dues to the discrete levels in the tip-induced QDot. c) Zoom at low bias on these conductance peaks. d) Pb NC grown on (110) surface of InAs. The red arrow indicates the line along which the spectra showed panels b) and c) have been measured. e) Schematic representation of the Pb NC on the InAs surface, below which an electron accumulation layer should exist. f) 1~$\mu$m$ \times $1~$\mu$m topographic image of Pb NCs accumulating along an atomic step edge of InAs. g) Zoom on the area indicated by a red arrow on panel f. h) Color map of the DC plotted as a function of sample bias and distance along the arrow indicated on panels f) and g). This color map displays large fluctuations in the energy of the discrete levels of the tip-induced QDot observed in the energy range [0-100 meV]. Near zero-bias, a small superconducting gap induced by proximity effect is observed. The orange curve is a DC curve taken at the position indicated by dash orange line.}
	\end{center}
\end{figure}

 Fig.~3bc shows the Differential Conductance (DC) dI/dV measured on the InAs surface at several distances, from 0 to 10 nm, of a Pb NC shown Fig.~3d. The data indicate that the Fermi level is in the conduction band as expected for this n-doped InAs sample. With a dopant concentration, $N_D\sim 6\times10^{16} cm^{-3}$, the Fermi level $E_F$ is 21 meV above the conduction-band minimum. A zoom of these spectra on the energy range [-150, 100] meV, Fig.~3c, shows the presence of small conductance peaks. These peaks result from tip-induced QDots, as already observed in past works \cite{Dombrowski1999-ia,Morgenstern2012-oa}. So far, these tip-induced QDots have been observed only on the (110) surface of InAs, confirming that the Fermi level is not pinned at this surface. We also find that these tip-induced QDots are present near the Pb NCs, which demonstrates that the surface states induced by Pb deposition on InAs are only localized below the Pb NC and do not extend significantly far from the NC, as sketched Fig.~3e. If the Pb-induced interface states were extending far from the NCs, the Fermi level would be pinned and no tip-induced QDots would be observed, in contradiction with the experimental observation. Fig.~3f shows a large area 1~$\mu$m$ \times $1~$\mu$m topographic image where NCs are observed to be accumulating along an atomic step edge of the InAs substrate. From this image, a zoom is extracted and shown Fig.~3g. Along the red arrow shown on this last panel, the DC is measured and shown as a distance-voltage map Fig.~3h. This map displays large variations in the energy of the tip-induced QDot, where the length scale for the shifts in these energy levels  is about 20 nm, which is of the order of the Fermi wavelength of the accumulation layer. These energy shifts likely result from the changing electrostatic environment as a consequence of the presence of the Pb NCs.

%Because of confinement effects, a discrete level structure is also expected below the metal with a maximum electron density located in the range 1-10 nm, sketched Fig.~1g, consistent with the estimated width of the tunnel barrier\cite{Tsui1970,Betti2001,Ando1982a}.  
\begin {table*}[ht!]

\begin{center}
\begin{tabular}{|c|c c c c c c c|}
	\hline  
	\\
	& NC Volume [nm$^3$] & NC Height [nm] & NC Area [nm$^2$] & E$_{\rm C}$ [meV] & $\delta_{\rm F1}$ [meV] & $\delta_{\rm F2}$ [meV]  & $E_{\rm Thouless}$ [meV] \\
	\\
	\hline
	\\
	
	 I & 807  & 5.5 & 324 & 14 & 0.2 & 0.14 & 44 \\ 
	 II & 627 &  5 & 278  & 15 & 0.3 & 0.18 & 48 \\ 
	 III & 275 & 3.8 & 172 & 35 & 0.6 & 0.4 & 63 \\
	 IV & 160 & 2.5 & 120 & 52 & 1.1 & 0.7 & 75 \\ 
	 V & 10  & 0.7 & 15 & 200 & 17 & 11 & 191 \\ 
	 VI & 1.5 & 0.4 & 5 & 1040 & 123 & 77 & 364 \\ 
    \\
	\hline

\end{tabular}
\caption {\textbf{NCs parameters}}
\end{center}
\end{table*}

\subsection{Coulomb Blockade}

\begin{figure}[ht!]
	\begin{center}
		\includegraphics[width=8cm]{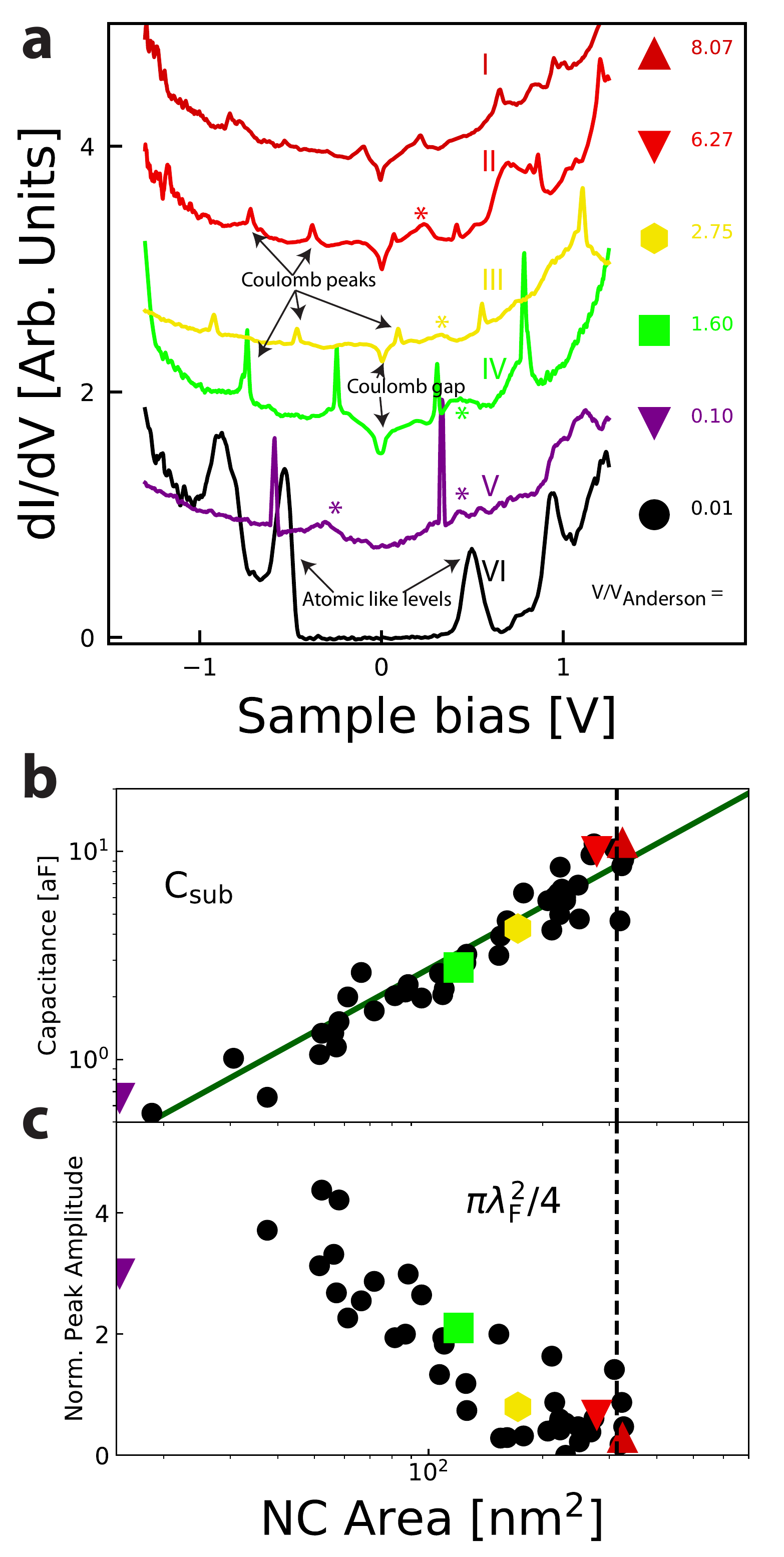}
		\caption{\label{Fig_Coulomb} a) DC spectra for the six NCs shown in Fig.~1. The spectra are indicated in order of increasing volume, in units of Anderson volume. For the largest NCs, I to V, Coulomb peaks and a zero-bias Coulomb gap are observed, as well as quantum well states indicated by stars. For the smallest Pb NC ($\simeq 0.01~$V$_{\rm Anderson} \simeq 1$~nm$^3$), the DC shows only broad atomic like levels separated by a large energy. No Coulomb blockade peaks are observed. b) Substrate-NC capacitance $C_{\rm sub}$ plotted as a function of NC area in contact with the substrate. c) Normalized amplitude of the Coulomb peaks as a function of the NC area. The peak amplitude goes to zero for NC area approaching $\pi\lambda_{\rm F}^2/4$ where $\lambda_{\rm F}$ is the Fermi wavelength of the 2D electron accumulation layer at the InAs surface.}
	\end{center}
\end{figure}

On NCs of six distinct sizes shown Fig.~1, representative DC spectra are shown Fig.~4a. They display a Coulomb gap at zero bias and sharp Coulomb peaks at higher voltage. Similar DC spectra with sharp Coulomb peaks have been observed previously\cite{McGreer1989-fe,Hanna1991-jb,Hanna1992-ic,Bar-Sadeh1996-hf,Hong2013-hm}. In Ref.~\cite{Vlaic2017-bs}, we show that the weak coupling model of of Ref.\cite{Hanna1991-jb} describes nicely the Coulomb characteristics observed in the IV curves. This weak coupling model describes four regimes, labelled I to IV, distinguished by the ratio $C_{\rm tip}/C_{\rm sub}$ and the fractional residual charge $Q_0$ on the NC. For our system Pb/InAs, the tip-NC capacitance is within the range $C_{\rm tip}\approx 0.1 - 0.5~$aF, while the substrate-NC capacitance is within the range $C_{\rm sub}\approx 1 -10~$aF. While both values are increasing with the NC area A, we find that the ratio $C_{\rm sub}/C_{\rm tip}\approx 10$ is only weakly changing. This value of the capacitance ratio implies that case III of Ref.\cite{Hanna1991-jb} applies to our system, where the residual charge has a negligible effect on the width of the Coulomb gap at zero bias. In this case, the width of the Coulomb gap at zero bias is given by $\delta V_{\rm sub}=e/(C_{\rm sub}+C_{\rm tip})$. The voltage interval between the peaks provides the addition voltage $\delta V_{\rm add}$ for an electron, which is related to the addition energy by : $\delta V_{\rm add}=E_{\rm add}/e\eta$ where $\eta=\frac{C_{\rm tip}}{C_{\rm tip}+C_{\rm sub}}\approx 0.1$ is the arm lever. 

Fig.~4b, extracted from Ref.~\cite{Vlaic2017-bs}, shows that $C_{\rm sub}$ extracted from the Coulomb gap at zero bias is a linear function of NC area A. From this dependence $C_{\rm sub}=A \varepsilon/d$, using $\varepsilon$ = 12.3, the dielectric constant of InAs one extracts $d$ = 4~nm for the effective tunnel barrier thickness. Table 1 shows a summary of parameters extracted for these NCs, i.e. the NC volume, the NC height,  the NC area, the Coulomb energy, the calculated mean level spacing for the two Fermi surfaces of Pb and the Thouless energy.

\begin{figure}[ht!]
	\begin{center}
		\includegraphics[width=7cm]{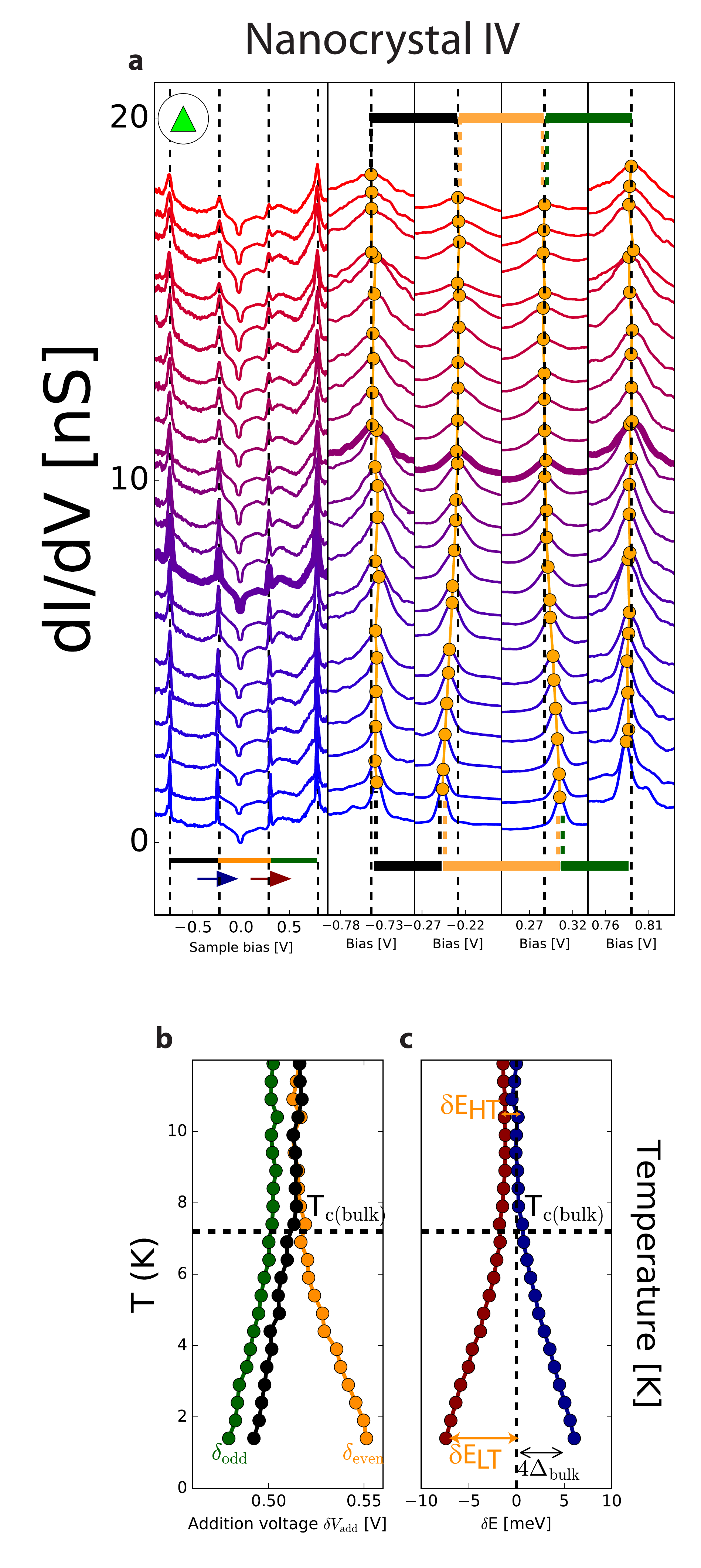}
		\caption{\label{Fig_parity} a) DC spectra as a function of temperature for NC IV with V/V$_{\rm Anderson}= 1.6$. The voltage separation between the Coulomb peaks, i.e. the addition voltage,  is indicated by the horizontal bars of different colors. In the same panels, zoom on the Coulomb peaks are shown where the maxima are indicated by orange dots. The DC spectra measured at T$_{\rm c}$ is plotted with a thicker linewidth than other spectra. b) The addition voltages as a function of temperature, where the color of the curves correspond to the horizontal bars indicated in panel a). c) The Difference in addition energies between two charge configurations given by  $\delta E=\eta(\delta V_{\rm Head}-\delta V_{\rm Tail})$, where the head (tail) refers to the arrows shown in panel a).  For panels b) and c), the value $T_c(\rm bulk)$ is indicated as horizontal dash lines. A double-headed arrow provides the scale for the energy gap $4\Delta_{\rm bulk}$ of bulk Pb.}
	\end{center}
\end{figure}

The addition energy, i.e. the energy for adding one electron into the NC, is given by :

\begin{equation}
\label{additionenergy}
E_{\rm {even (odd)}}=\frac{e^2}{C_{\small{\Sigma}}}+(-)2\Delta+\delta
\end{equation}
	
where the first term is the Coulomb energy, the second term depends on the parity of electron occupation number as a consequence of the formation of a Cooper pair\cite{Averin1992-lp,Lafarge1993-xf}, the third term is the electronic level spacing in the NC.

\begin{figure}[ht!]
	\begin{center}
		\includegraphics[width=7cm]{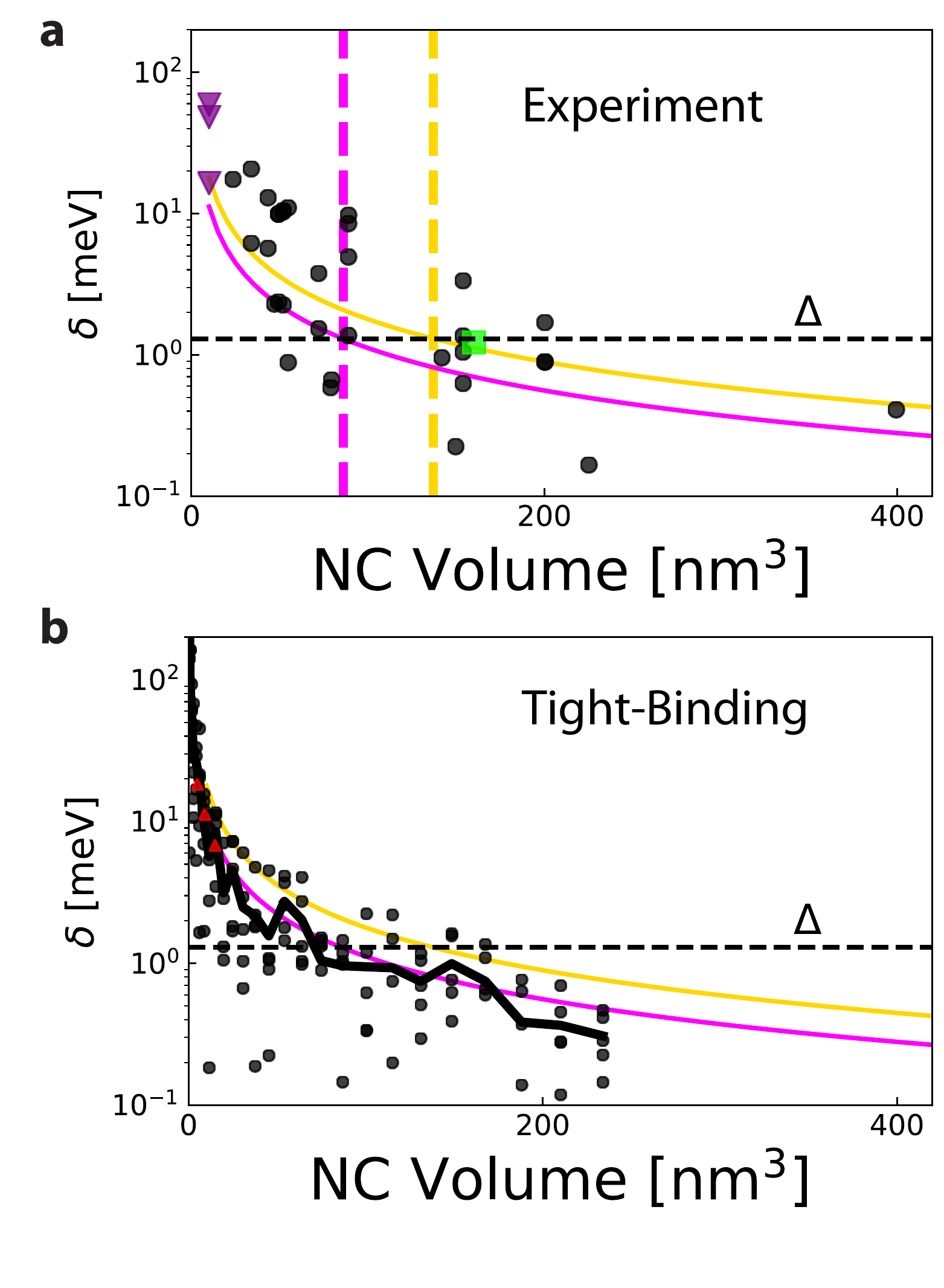}
		\caption{\label{Fig_Level_Spacing} a) Level spacing extracted from the addition energy measured above $T_{\rm c}$. The lines are the calculated mean level spacing, using relation~\ref{levelspacing}, for the two Fermi surfaces FS$_1$ (yellow) and FS$_2$ (pink) of Pb. b) The scattered symbols are the results from tight binding calculations of the level spacing for pyramidal NCs. The three red triangles are calculated level spacing from flat NCs. The black line is the average level spacing calculated from the scattered symbols. The colored lines are the calculated mean level spacings, using relation~\ref{levelspacing}, for the two Fermi surfaces of Pb.}
	\end{center}
\end{figure}

Thanks to the quality of the tunnel barrier enabling very sharp Coulomb charging peaks,  it has been possible to study, for the first time by STM, the evolution of the superconducting parity effect in the Coulomb charging energy across the Anderson limit\cite{Vlaic2017-bs}. As shown Fig.~5 for NC IV, in this experiment, the level spacing is obtained from the difference between successive addition energies $\delta=\delta E_{\rm HT}$ measured at high temperature ($T>T_{\rm c}$). From the difference between successive addition energies $\delta=\delta E_{\rm LT}$ at low temperature, the superconducting parity effect has been extracted. A detailed analysis of this parity effect as a function of NC volume is given in Ref.~\cite{Vlaic2017-bs}. 
A plot of the level spacing extracted from the addition energies at high temperature is shown Fig.~6a. The strong scattering observed in those data likely reflects the random structure of the spectrum where the level interval varies strongly.  Despite this scattering, one sees that the relation~\ref{levelspacing} for the \emph{mean} level spacing properly describes the evolution of the level spacing with decreasing NC volume. Fig. 6b shows the level spacing calculated from tight-binding calculations for pyramidal NCs as observed experimentally.
In these calculations, each Pb atom of a NC is characterized by two $s$ and six $p$ atomic orbitals, including the spin degree of freedom. The spin-orbit coupling is taken into account. The hopping terms are written in the two-center approximation and include interactions up to second-nearest-neighbors. The tight-binding parameters are given in Supp. Info.\cite{SuppInfo}  
The calculations can reproduce the average level spacing as a function of NC volume and, interestingly, the scattering in the level spacing.
For the sake of comparison with experiments, flat NCs made of 4 atomic rows were also considered for the smallest sizes. Interestingly, the average level spacing follows the same function of the NC volume as for pyramids.

\subsection{Nature of the tunnel barrier}

The observation of Coulomb blockade and quantum confinement in those Pb NCs raises the question of the nature of the tunnel barrier between the NCs and the substrate. As discussed above, no Schottky barrier is expected at metal-InAs interfaces\cite{Monch2001-ab,Morgenstern2012-oa} but instead an electron accumulation layer is expected in the InAs below the Pb NC. The origin of the tunnel barrier appears clearly after plotting the amplitude of the charging Coulomb peaks as a function of the NC area, Fig.~4c. One sees that the amplitude goes to zero for a NC area about $\pi\lambda_{\rm F}^2/4\simeq300$~nm$^2$, where $\lambda_{\rm F}=20$~nm is the Fermi wavelength of the electron accumulation layer below the Pb NC. This wavelength is calculated assuming that the Fermi energy $E_{\rm F}=150$~meV, of the accumulation layer is at the charge neutrality level\cite{Monch2001-ab,Morgenstern2012-oa}. As known from numerous works with quantum point-contacts formed in 2D electron gas\cite{Van_Wees1988-aj,Pasquier1993-pi}, the transmission coefficient $T$ decreases for constrictions smaller than the Fermi wavelength. Thus, in our experiment, because the area of NCs is smaller than $\simeq\lambda_{\rm F}^2$, their transmission coefficient with the 2D gas is significantly smaller than one, which corresponds to a tunneling regime and explains the observation of the Coulomb blockade. 

Thus, the dielectric thickness $d$ = 4 nm extracted from $C_{\rm sub}$ above is actually set by the Debye length of the 2D gas and $C_{\rm sub}$ actually corresponds to the quantum capacitance of InAs. Finally, a survey of the literature shows that Coulomb blockade has already been observed in metallic clusters deposited on InAs\cite{Wildoer1996-yl,Wiebe2003-ry,Muzychenko2009-rc}. The nature of the tunnel barrier was not identified in those works, though.

\section{Quantum Confinement}

In addition to the Coulomb peaks, the DC curves of Fig.~4a also display additional peaks resulting from discrete levels induced by the quantum confinement. On the largest NCs I to V, the peaks are of small amplitude, they are indicated by stars in Fig.~4a. On the smallest NC, labeled VI, no Coulomb peaks are observed but only large atomic-like levels resulting from strong quantum confinement in this atomic cluster. These three distinct regimes of quantum confinement are now discussed in more details.

\begin{figure}[ht!]
	\begin{center}
		\includegraphics[width=7cm]{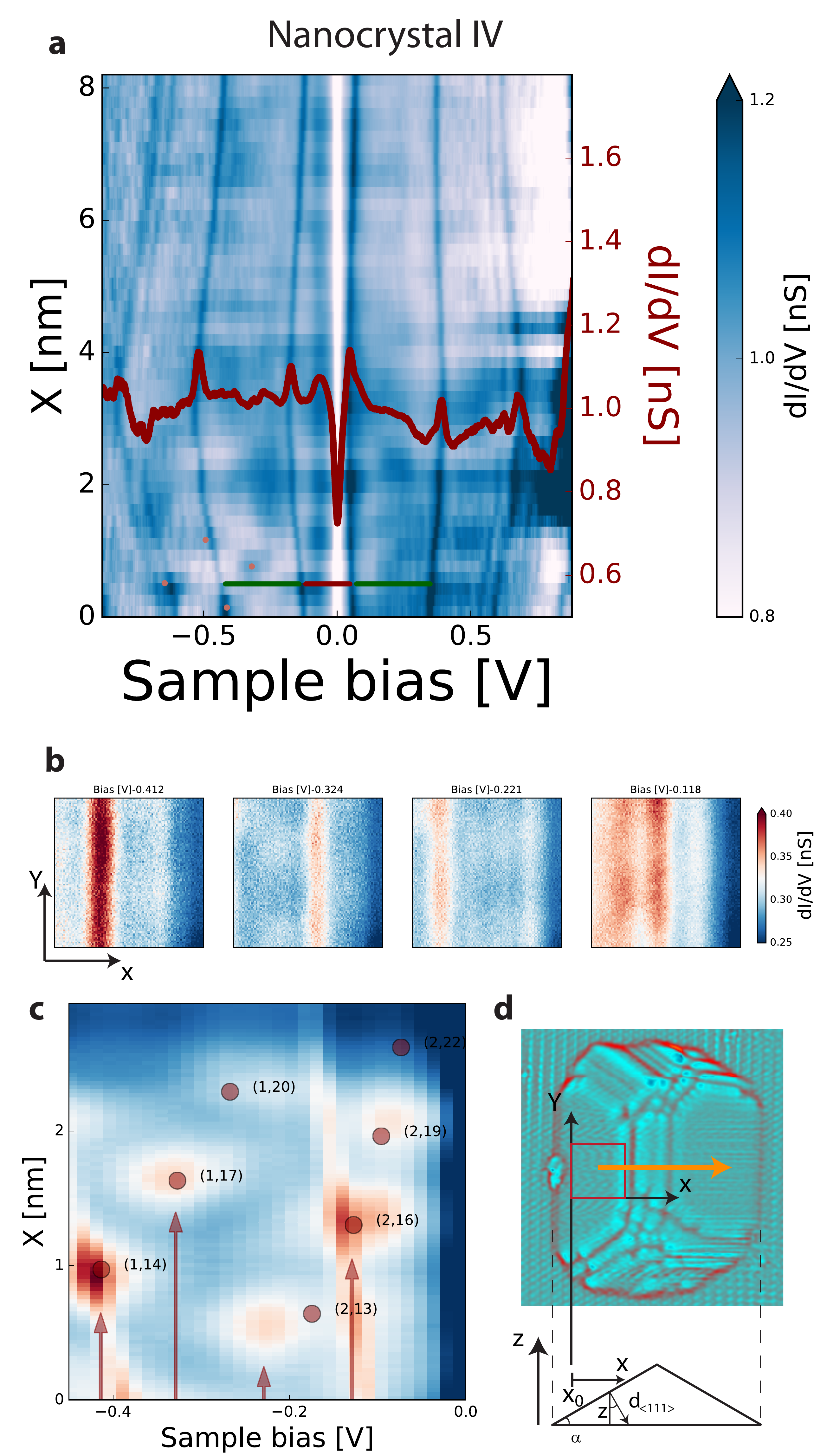}
		\caption{\label{Fig_Confinement} a) DC map as a function of sample bias and direction X measured along the arrow indicated in panel d. The map shows that the energy of the Coulomb peaks change with the tip position as a consequence of the changing tip-NC capacitance. They also show faint maxima indicated by small red dots. These maxima are seen more clearly on panel c. b) DC maps measured at different sample voltages on the X-Y area indicated by a red square on panel d. At these selected voltages, the QWSs appear as maxima of the differential conductance along the direction X. The voltage position of these maxima does not change along the Y direction. This shows that the energy of the QWSs at any (x,y) point on the NC depends only on the thickness of the NC, i.e. the length along the $\langle 111 \rangle$ direction as sketched panel d. Averaging the DC maps along the Y direction leads to an X-voltage map shown panel c. The vertical red arrows are located at the voltages of the maps shown in panel b. They indicate local maxima that correspond to the QWSs that appear as vertical lines in panel b. The red dots are the coordinates of the QWSs obtained by the phase accumulation model, see text, labeled by the index (P,n). The energy of the QWSs changes in the X direction following the change in the length d$_{<111>}$. These QWSs are also visible in the panel a, indicated by small red dots.}
	\end{center}
\end{figure}

To identify the origin of conductance peaks indicated by stars in Fig.~4a, Fig.~7a shows a DC map as a function of X-voltage on NC IV, along the arrow shown on the NC topographic image Fig.~7d. One can see: first, that the voltage position of Coulomb peaks changes slightly with the tip position above the NC, as a consequence of the variation in the tip-NC capacitance $C_{\rm tip}$; second, faint local conductance maxima indicated by red dots. Fig.~7b shows DC maps as a function of lateral XY position measured on the square area indicated by thin lines Fig.~7d. These maps are plotted only at selected bias indicated by red arrows on Fig.~7c. These maps show maxima running along the Y direction of the NC, which is the direction of constant  NC height as sketched Fig.~7d. This observation indicates that the observed maxima are the consequence of quantum confinement along the vertical direction Z of the NC. Averaging these maps along the Y direction leads to a DC map as a function of X-voltage, shown Fig.~7c, where one sees appearing clear local maxima as in Fig.~7a. These local maxima correspond to quantum well states whose energy is essentially controlled by the thickness of the NC. Similar quantum confinement has also been observed for $\langle 111 \rangle$ oriented 2D Pb thin films grown on silicon, where the quantum well states have been observed by photoemission\cite{Zhang2005-bi} and STM\cite{Altfeder1997-yq,Su2001-wp}.

The X-voltage coordinates of these states on the DC map Fig.~7c can be reproduced properly by the simple phase accumulation model\cite{Echenique1978-tl,Zhang2005-bi}. In this model, the condition for a standing wave and the formation of quantum well states is :

\begin{eqnarray}
2k(\varepsilon)Nd_{111}+\delta\phi=2\pi n
\end{eqnarray}

where N is the number of atomic layers traveled by the electrons in the $\langle 111 \rangle$ direction, n is an integer, and $\delta\phi$ is an additional phase shift experienced by the electron at the boundaries of the NC. The different families of quantum well states are labeled by the index P=2n-3N.
Thus, this Fabry-Perot regime of quantum confinement produces states regularly organized in space and energy and constitutes the regime of quantum confinement observed in large Pb NCs.

For very small NCs, i.e. atomic clusters, such as NC VI, of volume about $1.5~$nm$^3$, the DC presents large conductance peaks that result from quantum confinement in all directions of the NC. This spectrum has an atomic-like look with large energy level separation about almost 1~eV as visible on Fig.~4a. This atomic-like regime constitutes the regime of quantum confinement observed in atomic clusters.

\begin{figure}[ht!]
	\begin{center}
		\includegraphics[width=8cm]{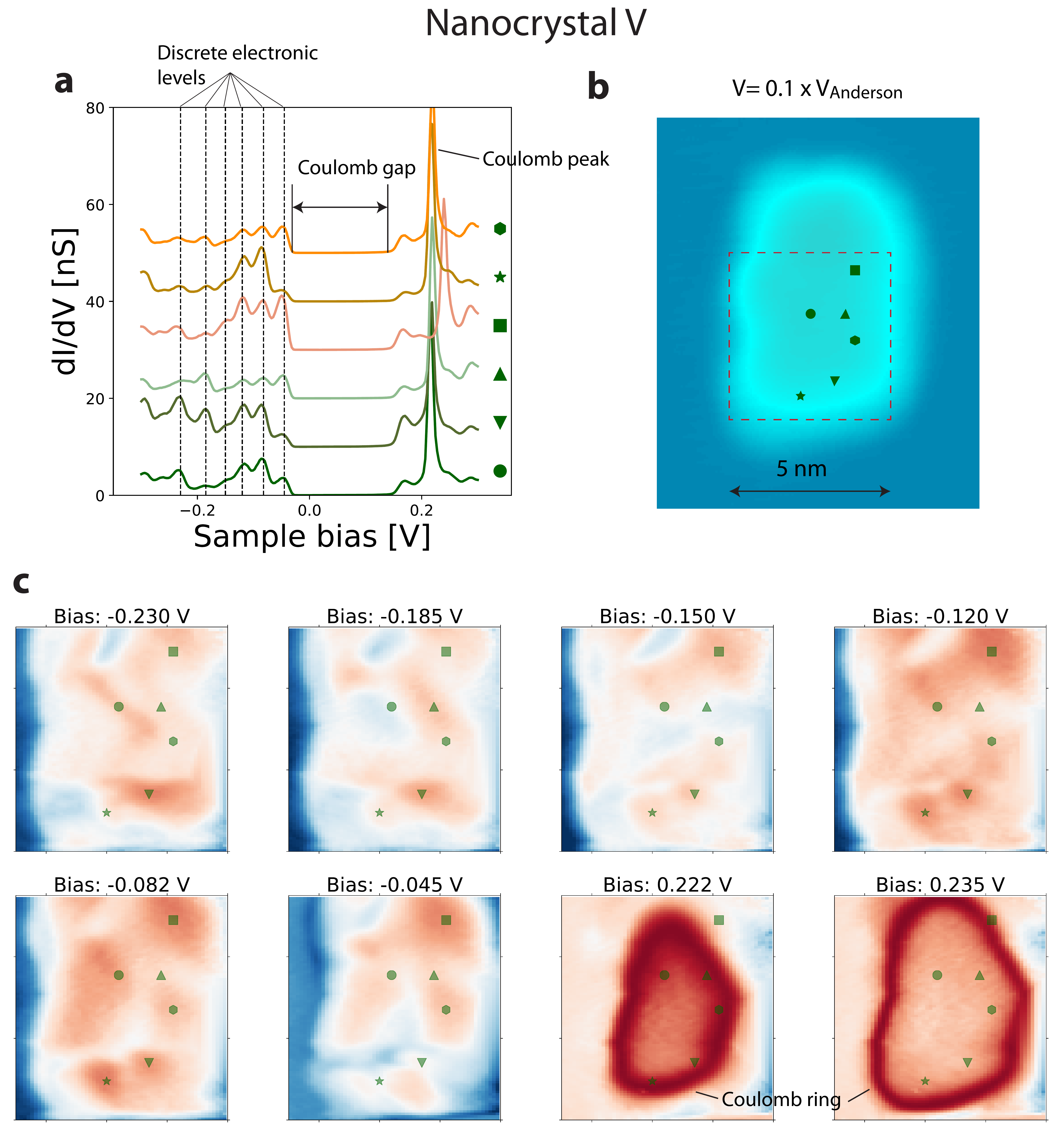}
		\caption{\label{Fig_confinement2} a) DC spectra as a function of sample bias measured at different positions indicated by symbols on the NC shown panel b). The spectra display a single Coulomb peak, a Coulomb gap and 6 discrete electronic levels. b) Topographic image of the NC. c) DC maps taken at the different voltages indicated by dash lines on panel a) on the XY area indicated by a dash red square on panel b). The Coulomb peak appears as a Coulomb ring on the DC maps taken at V$_{\rm Bias}=0.222$~V and V$_{\rm Bias}=0.235$~V.}
	\end{center}
\end{figure}

For intermediate NC volume about 10~nm$^3$ between the Fabry-Perot regime and the cluster regime, such as NC V, the level spacing becomes large enough for the discrete electronic levels to be seen as small peaks in the DC, as indicated by dashed lines on Fig~\ref{Fig_confinement2}a. For this NC, the Coulomb gap at zero bias is about 200~mV. The 6 DC spectra shown in this figure are extracted from a grid of 128 $\times$ 128 = 16384 spectra taken on the square indicated by dotted lines on the topographic image Fig.~\ref{Fig_confinement2}b. The location where these 6 spectra have been taken are indicated by symbols on the topographic map Fig.~\ref{Fig_confinement2}b and DC maps Fig.~\ref{Fig_confinement2}c. From this spectra grid, one can extract maps of the DC at 8 different voltage values, they are shown Fig.~\ref{Fig_confinement2}c. The first 6 DC maps are taken at voltage values corresponding to the discrete electronic levels, they show that the amplitude of these small peaks changes with the lateral position on the NC. The last two DC maps are taken at voltages values close to the Coulomb peak value, they show Coulomb rings that correspond to contours of constant electrostatic energy.  

\begin{figure}[ht!]
	\begin{center}
		\includegraphics[width=7cm]{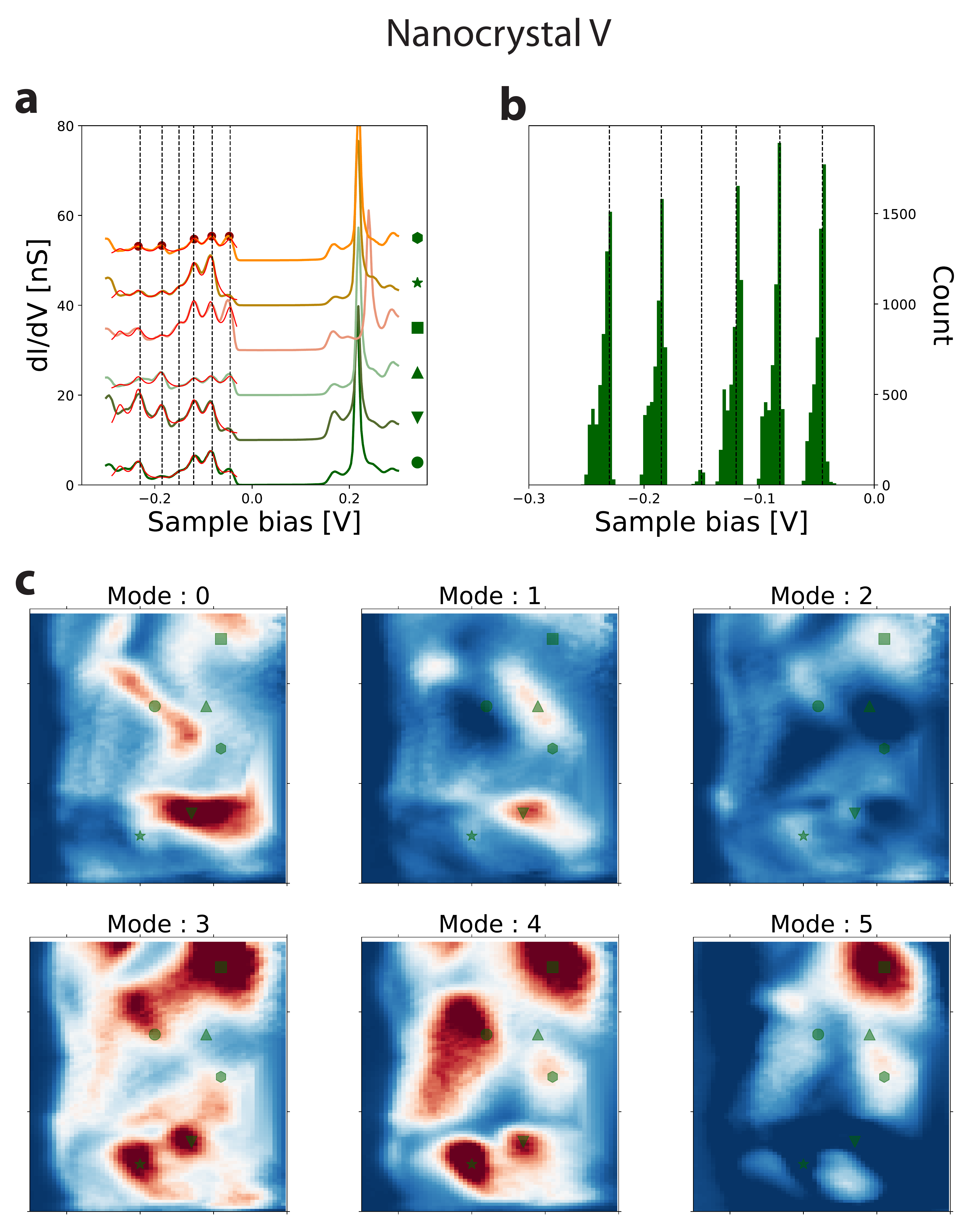}
		\caption{\label{Fig_confinement3} a) DC spectra identical to those shown Fig.~\ref{Fig_confinement2}a. The local maxima in the DC curve due to the discrete levels are identified by red dots. In the voltage range [-0.275~V,-0.025~V], the spectra can be fitted by the sum of six Lorentzian centered on the voltage values extracted from the histogram panel b. The fit are shown as thin red curves. The green symbols on the left indicate on the maps shown panels c the XY position where the spectra have been taken. b) Histogram of the voltage positions of the local maxima identified in the 128$\times$128 acquired DC spectra. The histogram shows only 6 well defined peaks indicating that only 6 discrete levels exist on this energy range. These 6 voltage values are used as the voltage positions of the Lorentz functions used to fit the DC spectra, as shown in panel a. c) Maps of the amplitude of the 6 Lorentzian as a function of position XY. These maps can be interpreted as maps of the amplitude of the wave-functions associated with the discrete levels.}
	\end{center}
\end{figure}

To quantify the number of discrete levels in the NC, we run an algorithm on all the 16384 spectra to find all local maxima on the voltage range [-0.275~V,-0.025~V], as shown Fig~\ref{Fig_confinement3}a where the local maxima are indicated by red dots. Then, a histogram of the voltage positions of the local maxima is plotted, Fig.~\ref{Fig_confinement3}b, and shows that there are only 6 well-defined peaks on this voltage range.

This observation of 6 peaks only into the histogram allow concluding that the electronic spectrum of the NC can be described, on this voltage range, by only 6 distinct electronic levels. From the voltage difference between the peaks, we find an average level spacing about 35~meV. This value is about 2 times larger than the theoretical value calculated for the Fermi surface FS$_1$ and about 3 times larger than the value calculated for the Fermi surface FS$_2$, see Table 1. For this small NC, an error on the volume by a factor of 2 cannot be excluded, as this corresponds to a change of $25\%$ for the linear size of the NC. Furthermore, shell effects or random level distribution effects could possibly explain this discrepancy between the measured level spacing and the expected theoretical mean value. Despite this discrepancy, the small peaks can be safely assigned to single electronic levels within the Pb NC.

To extract the amplitude of the wave-function for each state, we fit the spectrum on the voltage range [-0.275~V,-0.025~V] with the function~\ref{Eq:Lorentz}, i.e. the sum of 6 Lorentzian describing the 6 levels centered at the energies $\varepsilon_i$ identified in the histogram and linewidth $\Gamma=13$~meV :

\begin{equation}
\label{Eq:Lorentz}
\rho(V)=\sum_{i=0}^5{\frac{A_i}{(\varepsilon-\varepsilon_i)^2+\Gamma^2}}
\end{equation}

Following the fit of all spectra, we can plot maps of the amplitude A$_i$ of the Lorentzian as a function of the XY position, shown Fig~\ref{Fig_confinement3}c. These maps represent the amplitude of the wave-functions associated with the discrete electronic levels. 

While mapping wave-functions has already been done previously on InAs QDots \cite{Millo2001-ls}, where wave-functions of simple spherical or toroidal symmetry were observed, it is the first time that a map of the wave-functions of discrete levels in metallic NC is presented. In contrast to the QDot InAs, we find that the wave-functions in our metallic NCs have a random structure. This is actually not surprising. As shown Table 1, the Thouless energy for this NC is about 190 meV which is above the energy of these discrete levels, except for the first one. This implies that the states are in the chaotic regime and the wave-functions should have a random structure\cite{Alhassid2000-cd,Mirlin2000-xa}.

%While we do not observe enough discrete levels to attempt a study of the distribution of level spacing, the resolution and density of wave-function is sufficient for such a study. Fig~\ref{Fig_confinement3}d show histogram of the wave-function amplitude, which is compared to the Porter-Thomas distribution\cite{Alhassid2000-cd}: $P(t)=\frac{\exp{-t/2}}{\sqrt{2\pi t}}$. This distribution is expected to describe the distribution in the amplitude of the wave-function in the Gaussian ensemble, i.e. with time-reversal symmetry and no spin-orbit interactions. This last figure shows clearly that the experimental data deviates from this theoretical distribution. One obvious way to improve the theoretical modeling would be to include spin-orbit interactions. To our knowledge, the distribution of amplitude wave-function in the symplectic ensemble, which includes spin-orbit interactions, has not been established theoretically.

\section{Conclusion}
While STM seems an ideal experimental method for the study of quantum confinement effects in highly crystalline NCs grown in UHV, its development was hampered by two contradicting requirements: first, the substrate should be conducting enough to provide a current path to the ground, second, the NC should be separated from this substrate by a tunnel barrier. Using an InAs substrate presenting an electron accumulation layer of large wavelength $\sim 20$~nm, we found that NCs of lateral size smaller than this length are in the regime of Coulomb blockade. This results from the constriction of the electronic wave-function across this interface whose lateral size is smaller than the Fermi wavelength, implying that the electronic transmission across this interface drops below unity, even in the absence of any \emph{real} insulating barrier at the interface between the NC and the InAs substrate. This highly clean tunnel barrier enabled the first STM observation of the superconducting parity effect\cite{Vlaic2017-bs}, an unambiguous test of the Anderson criterion for the existence of superconductivity\cite{Vlaic2017-bs} and finally, in this manuscript, the observation of discrete electronic levels due to quantum confinement in the Pb NCs. We identified three regimes of quantum confinement. In the largest NCs, we found a Fabry-Perot regime where regular quantum well states are formed due to quantum interference along the $\langle 111 \rangle$ direction of the Pb NC. In the smallest NC, i.e. atomic clusters, we found atomic-like electronic levels. Finally, in the intermediate regime, we found the signature of discrete electronic levels in the DC spectra for which we mapped the corresponding wave-functions. We found that these wave-functions had a random spatial structure as expected for NCs in the chaotic regime of RMT. Future works in this direction with higher energy resolution at lower temperature may enable extracting correlations effects from the wave-function amplitude, such as due to Fermi statistics, superconducting correlations or more generally, electron-electron interactions. This observation of discrete electronic levels in metallic Pb NCs establishes today Pb/InAs as the most suitable system for the study of quantum confinement effects in metallic/superconducting NCs.

We acknowledge fruitful discussions with M. Aprili, C. Brun and B. Grandidier.

H.A. acknowledge support from ANR grant ‘QUANTICON’ 10-0409-01, China Scholarship Council and labex Matisse. D.R. and S.P. acknowledge C’NANO Ile-de-France, DIM NanoK, for the support of the Nanospecs project.

\bibliography{Bibliography}

\end{document}

\preprint{APS/123-QED}
\title{Supplementary Info : Quantum confinement effects in Pb Nanocrystals grown on InAs}
\author{Tianzhen Zhang}
\affiliation{LPEM, ESPCI Paris, PSL Research University, CNRS, Sorbonne Universit\'e, 75005 Paris, France}

\author{Sergio Vlaic}
\affiliation{LPEM, ESPCI Paris, PSL Research University, CNRS, Sorbonne Universit\'e, 75005 Paris, France}

\author{St\'ephane Pons}
\affiliation{LPEM, ESPCI Paris, PSL Research University, CNRS, Sorbonne Universit\'e, 75005 Paris, France}

\author{Guy Allan}
\affiliation{Univ. Lille, CNRS, Centrale Lille, ISEN, Univ. Valenciennes, UMR 8520 -
IEMN, F-59000 Lille, France}

\author{Christophe Delerue}
\affiliation{Univ. Lille, CNRS, Centrale Lille, ISEN, Univ. Valenciennes, UMR 8520 -
IEMN, F-59000 Lille, France}

\author{Alexandre Assouline}
\affiliation{LPEM, ESPCI Paris, PSL Research University, CNRS, Sorbonne Universit\'e, 75005 Paris, France}

\author{Alexandre Zimmers}
\affiliation{LPEM, ESPCI Paris, PSL Research University, CNRS, Sorbonne Universit\'e, 75005 Paris, France}

\author{Christophe David}
\affiliation{Centre de Nanosciences et de Nanotechnologies, CNRS, Univ. Paris-Sud, Universit\'es Paris-Saclay, C2N – Marcoussis, 91460 Marcoussis, France}

\author{Guillemin Rodary}
\affiliation{Centre de Nanosciences et de Nanotechnologies, CNRS, Univ. Paris-Sud, Universit\'es Paris-Saclay, C2N – Marcoussis, 91460 Marcoussis, France}

\author{Jean-Christophe Girard}
\affiliation{Centre de Nanosciences et de Nanotechnologies, CNRS, Univ. Paris-Sud, Universit\'es Paris-Saclay, C2N – Marcoussis, 91460 Marcoussis, France}

\author{Dimitri Roditchev}
\affiliation{LPEM, ESPCI Paris, PSL Research University, CNRS, Sorbonne Universit\'e, 75005 Paris, France}

\author{Herv\'e Aubin}
\email{Herve.Aubin@espci.fr} 
\affiliation{LPEM, ESPCI Paris, PSL Research University, CNRS, Sorbonne Universit\'e, 75005 Paris, France}

\date{\today}% It is always \today, today,
\maketitle

\section{Details on the crystallographic structure of the Nanocrystals}

The STM images enable a clear identification of the (111) planes of the Face Centered Cubic (FCC) structure, thanks to the hexagonal shape of the facets. The NCs have a pyramidal structure with truncated top. Two pyramidal shapes are possible with either an (100) basal plane or an (110) basal plane. As sketched Fig.~1, these two pyramidal shapes can be distinguished by the angle that the (111) plane makes with the basal plane. Experimentally, we find an angle close to  $\alpha \sim 35.3^{\circ}$, which corresponds to pyramids with a (110) basal plane.

\begin{figure}[ht!]
	\begin{center}
		\includegraphics[width=15cm]{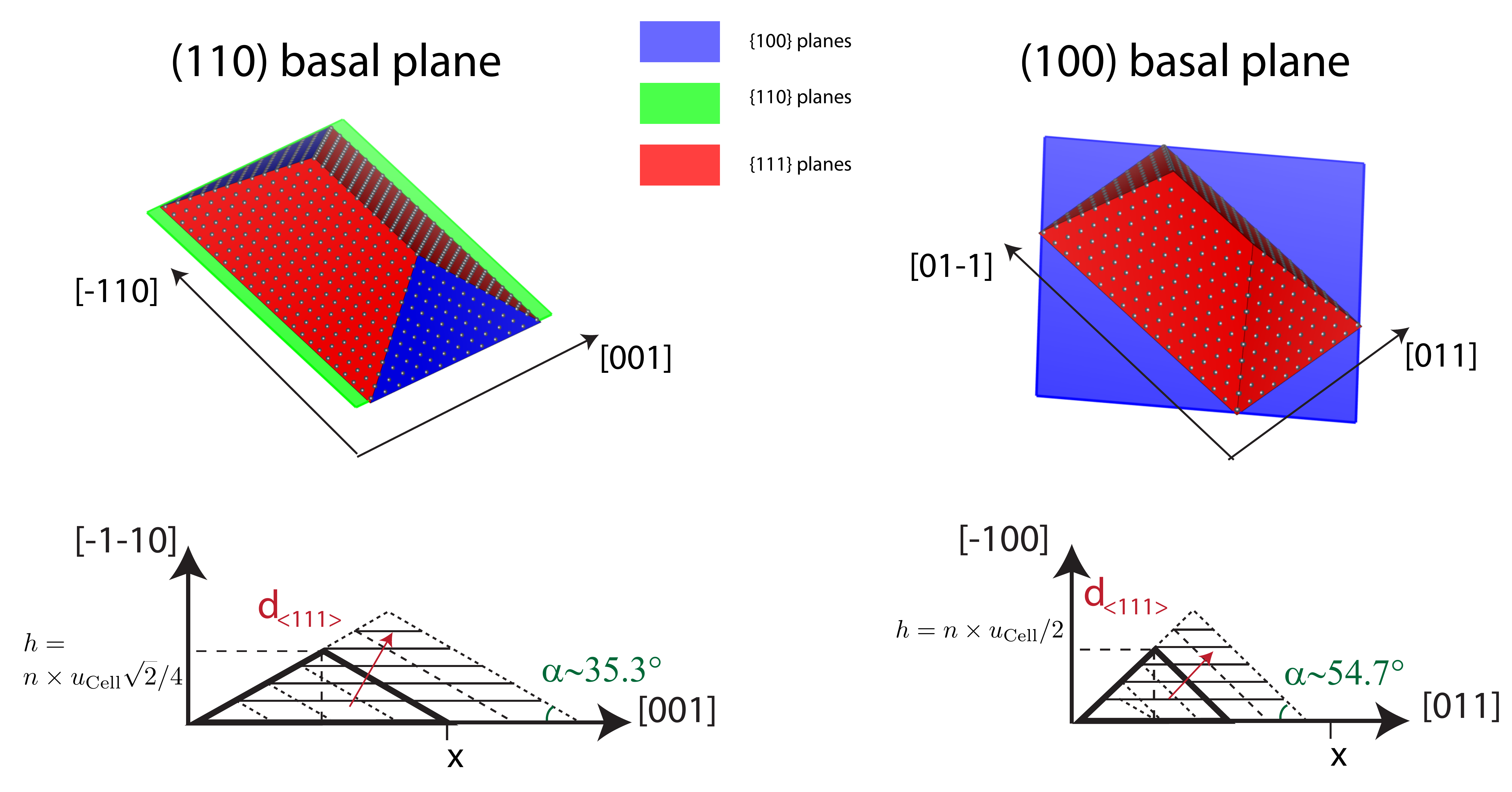}
		\caption{\label{Fig1} Sketch of the pyramidal NCs indicating the main crystallographic directions for (110) basal plane (left) and (100) basal plane (right).}
	\end{center}
\end{figure}

For these pyramids of base plane (100) and (110), the height of the nanocrystal is given by $h=n\times u_{\rm {cell}}/2$ and $h=n\times u_{\rm {cell}}\times\sqrt{2}/4$ respectively, where $u_{\rm {cell}}$=0.495 nm is the length of the unit cell of Pb and n is the number of atomic rows. The distance between atomic rows along the $\langle001\rangle$ directions is $d_{\langle001\rangle}=u_{\rm {cell}}/2= 0.247~$nm while the distance between atomic rows along the $d_{\langle011\rangle}=\langle011\rangle$ directions is $u_{\rm {cell}}\times\sqrt{2}/4=0.175~$nm.

The area of the base of the pyramid can be calculated from the relation $s=x^2$ with $x=2\times h/\tan{(\alpha)}$. The volume of the pyramid is $V1=s\times h/2$ for vertical end faces, while for a pyramid with triangular base, $V2=s\times h/3$. The expected theoretical heights and volumes plotted as function of NC base area are plotted Fig.~2 for the two possible pyramid structures of basal planes (100) and (110).
The base area of the NC is used as the varying parameter because the surface is the quantity measured with the highest precision by STM topography.

\begin{figure}[ht!]
	\begin{center}
		\includegraphics[width=15cm]{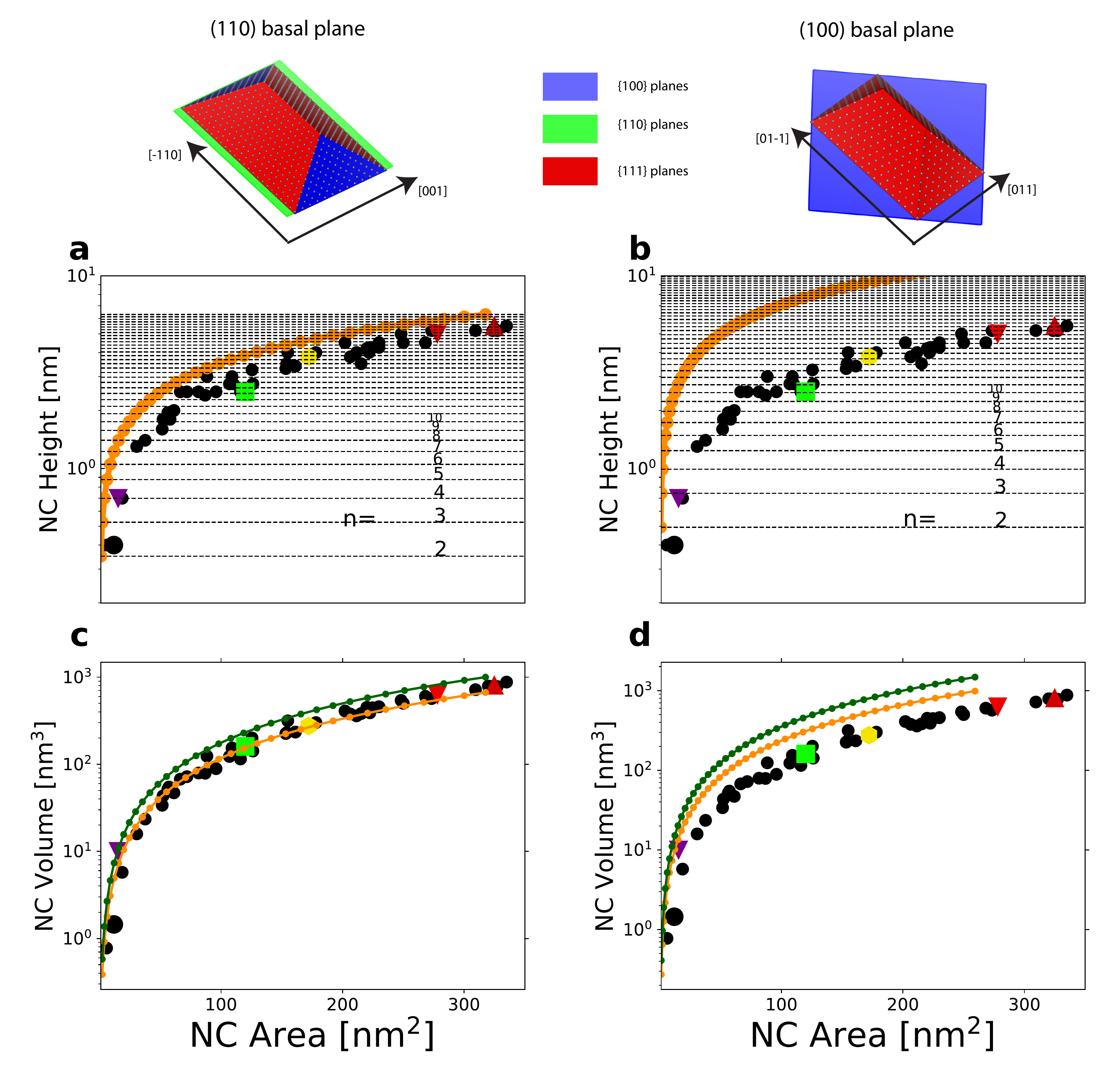}
		\caption{\label{Fig2} Theoretical height and volume of the NCs for the two possible pyramid structures with (110) basal plane (left) and (100) basal plane (right). They are compared to the experimental values extracted from the topographic STM images. }
	\end{center}
\end{figure}

The surface, the height and the volume of NCs are extracted from the STM topographic images and compared with the theoretical model Fig.~2. This plot shows that the pyramids with (110) basal plane best describe the data. As the area of the NC decreases, one sees that the height deviates significantly from the model which reflects that the pyramid have a truncated top. For the volume, the experimental value is obtained with good precision from a flooding method. For large NCs, the volume is properly describes by the theoretical model, deviations are getting larger in smalls NCs, again because of the truncated top.

\begin {table*}[ht!]

\begin{center}
\begin{tabular}{|c|c c c c c c|}
	\hline  
	\\
	& NC Height [nm] & NC Area [nm$^2$] & NC Vol [nm$^3$]  & n (rows) & $\delta_{\rm F1}$ [meV] & $\delta_{\rm F2}$ [meV] \\
	\\
	\hline
	\\
	 I & 5.5 & 324 & 807 & 31 & 0.2 & 0.14  \\ 
	 II &  5 & 278 & 627 & 29 & 0.3 & 0.18\\ 
	 III & 3.8 & 172 & 275 & 22 & 0.6 & 0.4 \\
	 IV & 2.5 & 120 & 160 & 14 & 1.1 & 0.7 \\ 
	 V & 0.7 & 15 & 10 & 4 & 17 & 11 \\
	 VI & 0.4 & 5 & 1.5 & 2 &123 & 77 \\ 
    \\
	\hline

\end{tabular}
\caption {NCs parameters (volume, height, surface) are extracted from the STM images. The effective number of atomic row is calculated from $n_{\rm eff}=h/d_{\langle011\rangle}$. The level splittings are calculated from the usual formula Eq.1 of main text.}

\end{center}
\end{table*}

Table~1 provides the experimental parameters (height, surface, volume) extracted from topographical STM images for selected nanocrystals I to VI. This table also provide the effective number of atomic rows calculated from $n_{\rm eff}=h/d_{\langle011\rangle}$.

\begin{table*}[ht!]

\begin{center}
\begin{tabular}{|c|c c c c c|}
	\hline  
	\\
	& n (Atomic rows) & NC Height [nm] &  NC Area [nm$^2$] & NC Vol (sh/2) [nm$^3$] & NC Vol (sh/3) [nm$^3$] \\
	\\
	\hline
	\\
	
	 I & 36  & 6.3 & 318 & 1000  & 667  \\ 
	 II & 34 &  5.95 & 283  & 843 & 562\\ 
	 III & 27 & 4.7 & 178 & 422 & 281 \\
	 IV & 22 & 3.85 & 119 & 228 & 152 \\ 
	 V & 8 & 1.4 & 15 & 11 & 7.3 \\ 
	 VI & 4 & 0.7 & 4 & 1.4 & 1 \\ 
    \\
	\hline

\end{tabular}
\caption {NCs parameters (n atomic rows) and corresponding height surface and volume calculated assuming a pyramid with (110) basal plane. The value n is adjusted such that the surface of the NC match the experimental value.}

\end{center}
\end{table*}

Table~2 provides best estimations of the pyramid parameters for NC I to VI obtained by adjusting the surface. One sees that this simple pyramidal model overestimates the number of atomic rows for smaller NCs, as a consequence of the truncated top. Consequently, the smallest NCs V and VI are best described as flat atomic layers of 4  and 2 atomic rows respectively.

\section{TB parameters of Lead (Pb)}

Knowing the detailled shape of the NCs, tight binding model is used to calculate the electronic levels spectrum. The parameters are taken from Ref.~\cite{Slater1954-nv} and given in table 3.

\begin{table}[h]
\begin{center}
  \begin{tabular}{llll}
    \hline
    \hline
    \multicolumn{4}{c}{Parameters for Pb (eV)} \\
    \hline
    $E_s$ & 3.52800 & $E_p$ & 13.59901 \\
    $\Delta$ & 1.14900 \\
    $V_{ss \sigma}({\rm 1})$ & -0.30900 & $V_{ss \sigma}({\rm 2})$ & 0.03000 \\
    $V_{sp \sigma}({\rm 1})$ & +0.47339 & $V_{sp \sigma}({\rm 2})$ & -0.01000 \\
    $V_{pp \sigma}({\rm 1})$ & +1.34301 & $V_{pp \sigma}({\rm 2})$ & +0.14600 \\
    $V_{pp \pi}({\rm 1})$ & -0.01278 & $V_{pp \pi}({\rm 2})$ & -0.00200 \\

  \end{tabular}
  
\caption{TB parameters (notations of Slater and Koster \cite{Slater1954-nv}) for Pb in an orthogonal $sp^3$ model. $\Delta$ is the spin-orbit coupling. (1) and (2) denote first and second nearest-neighbor interactions, respectively. Lattice parameter: $a= 4.95$ \AA.}
\label{table_param}

\end{center}
\end{table}
\bibliography{Bibliography}